\documentstyle[12pt,a4,epsf,subfigure,citesort]{article}

\def\e{\epsilon}
\def\ee{\e^2}

\def\aa{\alpha_1}
\def\ab{\alpha_2}
\def\aj{\alpha_j}

\def\k{\kappa}

\def\jab{,\ j = 1,2}
\def\kpaa{,\ k\ge 0}

\def\lpb{,\ l\ge 2}
\def\lab{,\ l=0,1,\cdots}

\def\pa{p_1}
\def\pb{p_2}
\def\pj{p_j}

\def\qa{q_1}
\def\qb{q_2}
\def\qj{q_j}

\def\ddqj{\Delta^2_\e q_j(t)}

\def\qan{q_{1,n}}
\def\qbn{q_{2,n}}
\def\qjn{q_{j,n}}
\def\qkn{q_{k,n}}

\def\qjnn{q_{j,n+1}}

\def\qjnpm{q_{j,n\pm 1}}

\def\pjn{p_{j,n}}

\def\pjnn{p_{j,n+1}}

\def\qjt{q_j(t)}
\def\qkt{q_k(t)}

\def\qjtn{q_j(t+\e )}

\def\qjtp{q_j(t-\e )}

\def\qjtpm{q_j(t\pm\e )}

\def\qjtt{\frac{d^2\qjt}{dt^2}}

\def\qjtea{\hat{q}_{j,0}(t,\e )}
\def\qktea{\hat{q}_{k,0}(t,\e )}

\def\qaute{q_{1,u} (t,\e )}
\def\qbute{q_{2,u} (t,\e )}
\def\qjute{q_{j,u} (t,\e )}

\def\qjste{q_{j,s} (t,\e )}

\def\qjumte{q_{j,u} (-t,\e )}

\def\ta{t_1}
\def\tb{t-2}
\def\tj{t_j}

\def\qjak{q_{j,0k}(t)}

\def\qjal{q_{j,0l}(t)}

\def\qjaa{q_{j,00}(t)}

\def\qjab{q_{j,01}(t)}

\def\qjaatt{\ds\frac{d^2\qjaa}{dt^2}}

\def\qjaltt{\ds\frac{d^2\qjal}{dt^2}}

\def\Fjak{F_{j,0k}}

\def\Fjal{F_{j,0l}}

\def\va{v_1(t)}
\def\vb{v_2(t)}

\def\qjtba{q_{j,10}(t)}

\def\ta{t_1}
\def\tb{t_2}

\def\ajk{a_{j,k}}
\def\ajl{a_{j,l}}

\def\aal{a_{1.l}}
\def\abl{a_{2,l}}

\def\aaa{a_{1,0}}
\def\aba{a_{2,0}}

\def\aab{a_{1,1}}
\def\abb{a_{2,1}}

\def\Pj{\Phi_j}
\def\Pk{\Phi_k}

\def\Plh{\hat{\Phi}_l}

\def\Pjze{\Pj (z,\e )}
\def\Pkze{\Pk (z,\e )}

\def\Pjz{\Pj (z)}

\def\Pz{\Phi (z)}

\def\Puz{\Phi_u(z)}
\def\Psz{\Phi_s(z)}

\def\Paazh{\hat{\Phi}_{0}(z)}
\def\Pabzh{\hat{\Phi}_{1}(z)}

\def\Palzh{\hat{\Phi}_{l}(z)}

\def\al{a_l}
\def\aal{a_{1,l}}
\def\abl{a_{2,l}}

\def\V{V(p)}

\def\Ppm{\Phi_{\pm}(z)}
\def\Pmz{\Phi_{-}(z)}

\def\Pm{\Phi_{-}}

\def\zb{\bar{z}}

\def\Pbar{\bar{\Phi}}

\def\Ps{\Phi_s}
\def\Pu{\Phi_u}

\def\Pc{\stackrel{\sss\circ}{\ds\Phi}}

\def\Pjb{\hat{\Phi}_{j,1}(z)}

\def\Pbe{\hat{\Phi}_1^{even}(z)}
\def\Pbo{\hat{\Phi}_1^{odd}(z)}

\def\dk{d_k}

\def\qjteb{\hat{q}_{j,1}(t,\e )}

\def\qjtec{\hat{q}_{j,2}(t,\e )}

\def\qjtel{\hat{q}_{j,l}(t,\e )}

\def\qjteci{\stackrel{\sss\circ}{q_j}(t,\e )}

\def\qjteut{\tilde{q}_{j,u}(t,\e )}
\def\qjteumt{\tilde{q}_{j,u}(-t,\e )}
\def\qjtest{\tilde{q}_{j,s}(t,\e )}

\def\pjteut{\tilde{p}_{j,u}(t,\e )}

\def\pjtest{\tilde{p}_{j,s}(t,\e )}

\def\qjtmeut{\tilde{q}_{j,u}(t-\e ,\e )}

\def\a{\alpha}
\def\b{\beta}

\def\d{\delta}
\def\k{\kappa}

\def\e{\epsilon}

\def\D{\Delta}

\def\ee{\e^2}

\def\aa{\alpha_1}
\def\ab{\alpha_2}
\def\aj{\alpha_j}

\def\DD{\D^2}

\def\Pj{\Phi_j}

\def\DDe{\DD_\e}

\def\sss{\scriptscriptstyle}

\def\ds{\displaystyle}

\def\Henon{H$\acute{\rm e}$non }

\def\N{{\bf N}}
\def\Z{{\bf Z}}
\def\R{{\bf R}}

\def\iR{i{\bf R}}

\def\D{\Delta}
\def\DD{\D^2}

\def\DDe{\DD_\e}

\def\sech#1{{\rm sech}(#1)}
\def\sechd#1{{\rm sech}^2(#1)}
\def\secht#1{{\rm sech}^3(#1)}

\def\sinh#1{{\rm sinh}(#1)}
\def\sinhd#1{{\rm sinh}^2(#1)}
\def\sinht#1{{\rm sinh}^3(#1)}

\def\cosh#1{{\rm cosh}(#1)}

\def\O#1{O(#1)}
\def\Osize#1{O\left(#1\right)}

\def\Re#1{{\rm Re\ }#1}
\def\Im#1{{\rm Im\ }#1}
\def\arg#1{{\rm arg}#1}
\def\Res#1#2{{\rm Res}[#1;#2]}

\def\Lap{L}
\def\LapIm{\Lap^{-1}}

\def\exp#1{e^{#1}}
\def\Exp#1{{\rm exp}\left(#1\right)}

\def\cosn#1{{\rm cos}#1}
\def\sinn#1{{\rm sin}#1}

\def\defeq{\stackrel{\rm def}{=}}

\def\const{{\rm const}}

\makeatletter
\def\@begintheorem#1#2{\rm \trivlist
     \item[\hskip \labelsep{\bf #1\ #2}] \item[]}
\def\@opargbegintheorem#1#2#3{\rm \trivlist
     \item[\hskip \labelsep{\bf #1\ #2\ (#3)}] \item[]}
\makeatother

\begin{document}

\title{Asymptotic expansion of 1-dimensional sub-manifolds of stable
  and unstable manifolds in a 4-dimensional symplectic mapping}

\author{Yoshihiro Hirata\thanks{e-mail address:
    yhirata@allegro.phys.nagoya-u.ac.jp}\, and Tetsuro Konishi\\
  Department of Physics, Faculty of Science, Nagoya University,\\
  Furo-cho, Chikusa-ku, Nagoya, 464-8602, Japan}

\begin{titlepage}

\begin{flushright}
  DPNU--98--45\\
  December 1998
\end{flushright}

\LARGE
\begin{center}
  Asymptotic expansion of 1-dimensional sub-manifolds of stable
  and unstable manifolds in a 4-dimensional symplectic mapping
\end{center}

\large
\begin{center}
  Yoshihiro Hirata\footnote{e-mail address:
    yhirata@allegro.phys.nagoya-u.ac.jp}\, and Tetsuro Konishi\\
  Department of Physics, Faculty of Science, Nagoya University,\\
  Furo-cho, Chikusa-ku, Nagoya, 464-8602, Japan
\end{center}
\vspace{10em}

\abstract{We analytically compute asymptotic expansions of a 1-dimensional
sub-manifold of stable and unstable manifolds in a 4-dimensional
symplectic mapping by using the method called {\it asymptotic
expansions beyond all orders}.
This method enables us to capture exponentially small splitting of
separatrices and also to obtain explicit functional approximations of
the sub-manifolds.
In addition, we show the condition with which homoclinic structure
caused by crossing between the stable and unstable sub-manifolds is
regarded as a direct product of 2-dimensional mappings.
}

\end{titlepage}

\normalsize

\section{Introduction}
\label{sec:Introduction}

Existence of transversal homoclinic (or heteroclinic) points implies
non-integrability of Hamiltonian systems and it is important to
understand homoclinic structure for studying phase space's structure
of Hamiltonian systems.
Melnikov's method (or Melnikov's integral) is a powerful tool to
search for transversal homoclinic points\cite{GH}.
This method is based on the regular perturbation method with respect
to perturbation strength $\e$, hence it is difficult to apply this
method to singular perturbed problems straightforwardly.

However in Hamiltonian systems with rapidly forces, and 2-dimensional
symplectic mappings which are perturbations from the identical
mappings, and so on, it is known that the homoclinic structure
contains singular terms, for example, $\Exp{-\beta/\e^\a}, \a, \b >
0$~\cite{Nei84,LST,Kadanoff}, and concequently had not been captured
with the regular perturbation method with respect to $\e$.

The difficulty however have been overcome by using the method called
{\it asymptotic expansions beyond all orders}.
This method was first proposed and applied to the standard mapping by
Lazutkin and co-workers\cite{GLS}. 
They calculated intersecting angles between stable and unstable
manifolds asymptotically.
And the method was improved by using Borel transformation and Stokes
phenomenon\cite{HM,TovinvLap}.
The same approaches were developed independently by Kruskal and
Segur\cite{KS} in the model of crystal growth and by Pomeau et
al.\cite{Pomeau} in the K-dV solitons under a 5-th order singular
perturbation. 
Afterwards Tovbis et al. applied this method to \Henon
map\cite{Tov1,Tov2} and constructed explicit functional approximations
of the stable and unstable manifolds.
Nakamura and Hamada analyzed a 2-dimensional symplectic mapping
with a double well potential in the same way and calculated
intersecting angles of stable and unstable manifolds\cite{NakaHama}.

The applications mentioned above are all to 2-dimensional symplectic
mappings, which correspond to Poincar\'e mappings of Hamiltonian
systems with two degrees of freedom.
On the other hand, 4-dimensional symplectic mappings corresponds to
Poincar\'e mappings of systems with three degrees of freedom.
It is very impotant to extend the analytical method for 2-dimensional
symplectic mappings to 4- or more-dimensional symplectic mappings,
because phase space's structure of Hamiltonian systems with three or
more degrees of freedom quite differs from that of systems with two
degrees of freedom, e.g., Arnold diffusion\cite{ArnoldDiffusion}.

An application to 4-dimensional symplectic mappings is reported by
Gelfreich and Sharomov\cite{GS95}.
They treat a coupled 4-dimensional standard map and prove the
splitting angle between 1-dimensional stable and unstable
sub-manifolds is exponentially small and the splitting angle
orthogonal to the sub-manifolds is not exponentially small.
Their model however contains the particular coupling potential, i.e.,
two-body central force coupling $\tilde{J}(q_1-q_2)$ (compare the
mapping (\ref{eqn:map0}) or (\ref{eqn:map1})).

We would like to take more general 4-dimensional mappings into
consideration and construct functional approximations of 1-dimensional
sub-manifolds of stable and unstable manifolds.
In this paper we shall show that it is possible to apply this method
to more general coupling potential terms.

In this paper we consider a 4-dimensional symplectic map
\begin{eqnarray}
  \label{eqn:map0}
  \left\{
    \begin{array}{rcl}
      \pjnn & = & \ds \pjn - \e \left( 2\qjn^3 - \qjn \right) -
      \e^{\gamma + 1}\k\frac{\partial\tilde{J}}{\partial\qjn} \\
      \qjnn & = & \qjn + \e\pjnn
    \end{array}
  \right.\jab,
\end{eqnarray}
where $0<\e\ll 1$, $\k\in\R$, $\gamma\in\N$, and $\tilde{J} =
\qan^{\a_1}\qbn^{\a_2},\, \a_1, \a_2 \in\N$ is a coupling term.
The map (\ref{eqn:map0}) is an {\it exact symplectic map}, hence there 
exists a generating function $W = I + \e J$~\cite{DR97}, where
\begin{eqnarray}
  \label{eqn:genfun}
  I & = & \pjnn\qjn,\\
  J & = & \sum_{j=1}^{2}{\left\{T(\pjnn) + V(\qjn)\right\}} +
  \e^\gamma\k\tilde{J}(\qjn),\\
  T(p) & = & \frac{1}{2}p^2,\\
  \label{eqn:potential}
  V(q) & = & \frac{1}{2}q^4 - \frac{1}{2}q^2.
\end{eqnarray}
And the map (\ref{eqn:map0}) is defined as $\qjnn = \partial W /
\partial \pjnn$ and $\pjn = \partial W / \partial \qjn$.
For $\e = 0$ the map is obviously identical, hence the map
(\ref{eqn:map0}) can be regarded as a perturbation from the
4-dimensional identical map.
Note that the map (\ref{eqn:map0}) becomes chaotic when $\e \ne 0$, in 
general.

The symplectic map (\ref{eqn:map0}) has a fully hyperbolic fixed
point, which is a direct product of two of 2-dimensional hyperbolic
fixed points, at the origin $(0,0,0,0)$ (see the eqnation
(\ref{eqn:potential})).
Hence there exist 2-dimensional stable and unstable manifolds near the
origin.

In the present work we take $\a_1 = \a_2 = 2$, $\gamma = 2$,
i.e., $\tilde{J} = \qan^2\qbn^2$ to simplify the following computation 
and notation.
Thus the symplectic map (\ref{eqn:map0}) is rewritten as
\begin{eqnarray}
  \label{eqn:map1}
  \left\{
    \begin{array}{rcl}
      \pjnn & = & \pjn - \e \left( 2\qjn^3 - \qjn \right) -
      2\e^3\k\qjn\qkn^2 \\
      \qjnn & = & \qjn + \e\pjnn
    \end{array},
  \right.j,k \in \{1,2\},\,j \ne k.
\end{eqnarray}

To construct the unstable manifold we transform the map
(\ref{eqn:map1}) into the second order difference equations and change 
the independent variable $n$ into $t$ with meanings of time.
We take $\e$ as the difference parameter, that is, $\qjnpm =
\qjtpm\jab$ if $\qjn = \qjt$~\cite{GLS,NakaHama,Tov1,Tov2}.
Thus
\begin{eqnarray}
  \label{eqn:mapqt}
  \ddqj & = & \qjt - 2\qjt^3 - 2\e^2\k\qjt\qkt^2
\end{eqnarray}
where $\ddqj = \{ \qjtn - 2\qjt + \qjtp \}/\e^2\jab$.

One can transform the equations (\ref{eqn:mapqt}) into
singular-perturbed ODEs\cite{Tov1,Tov2,NakaHama},
\begin{eqnarray}
  \label{eqn:outereqIntro}
  \ds \qjtt & = & \ds \qjt - 2\qjt^3 - 2\ee\k\qjt\qkt^2 -
  2\sum_{l=2}^{\infty}
  \frac{\e^{2(l-1)}}{(2l)!}\frac{d^{2l}\qjt}{dt^{2l}}
\end{eqnarray}
One of the purposes of this paper is to construct explicit functional
approximasions of 1-dimensional stable and unstable sub-manifolds
which satisfy $q_1(t,\e) = q_2(t,\e)$.

An outline of constructing functional approximations of the
sub-manifold, which is also an outline of this paper, is as
follows:
We construct solutions to the equations (\ref{eqn:outereqIntro}) with
the regular perturbation method (section \ref{sec:outer}).
The solutions however break down around $t \sim 0$ because of
existence of singular points.
Hence we blow up complex $t$-plane near one of the singular points and
analyze with singular perturbation methods (section \ref{sec:break}).
One can rewrite the equations (\ref{eqn:mapqt}) near the singular
points (section \ref{sec:Inner}) as
\begin{eqnarray}
  \label{eqn:Introinnereq}
  \Delta^2 \Pj & = & -2\Pj^3 + \ee\Pj - 2\k\ee\Pj\Pk^2,
\end{eqnarray}
where $\Delta^2 \Pjz = \Pj (z+1) - 2\Pjz + \Pj (z-1)$, and $j,k \in
\{1,2\}, j\ne k$.
We analyze the equations (\ref{eqn:Introinnereq}) by using Borel
transformation, which enables us to capture Stokes Phenomenon (section 
\ref{sec:InnerFormal}).
We match the solutions of the equations (\ref{eqn:Introinnereq}) to
the solutions of the equations (\ref{eqn:outereqIntro}) and
successfully obtain the analytical representation of a 1-dimensional
sub-manifold of an unstable manifold in the 4-dimensional symplectic
mapping (section \ref{sec:Matching}).
Finally we will conclude the result (section \ref{sec:result}) and
give a summary of this paper in section \ref{sec:Summary}.

\section{The outer equation}
\label{sec:outer}

In this section we try to construct a 1-dimensional sub-manifold of an
unstable manifold of the equations (\ref{eqn:map1}) by using the
regular perturbation method, i.e., Taylor expansion with respect to
$\e$.
Although the solution which satisfies the boundary conditions is
obtained, it does not converge.
We must therefore treat the equations (\ref{eqn:Introinnereq}) with
the singular perturbation method in the next and following sections.
Outline of construction of the manifold, which corresponds to contents 
from section \ref{sec:outer} to section \ref{sec:Matching}, goes along
the way by Tovbis et al.\cite{Tov1,Tov2}.
From now on, the subscripts $j$ and $k$ will be used as the meaning of 
$j,k \in \{1,2\}, j \ne k$ without further comment.

First using Taylor expansions as
\begin{eqnarray}
  \label{eqn:taylor}
  \qj(t\pm\e) & = & \sum_{l=0}^{\infty}\frac{(\pm\e
    )^l}{l!}\frac{d^l\qjt}{dt^l},\nonumber
\end{eqnarray}
one can rewrite the equations (\ref{eqn:mapqt}) as
\begin{eqnarray}
  \label{eqn:outereq}
  \qjtt & = & \qjt - 2\qjt^3 - 2\ee\k\qjt\qkt^2 - 2\sum_{l=2}^{\infty}
  \frac{\e^{2(l-1)}}{(2l)!} \frac{d^{2l}\qjt}{dt^{2l}}.
\end{eqnarray}
Equations (\ref{eqn:outereq}) and solutions to them are called {\it
  the outer equations} and {\it the outer solutions}, respectively.
One can construct asymptotic representations of an unstable manifold
as $t\to -\infty$ by solving the ODEs (\ref{eqn:outereq}).
Boundary conditions with which we solve the equations
(\ref{eqn:outereq}) are
\begin{eqnarray}
  \label{eqn:cond1}
  \left.
    \begin{array}{rcl}
      \ds \lim_{t\to -\infty}\qjute & = & 0\\
      \qjute & > & 0
    \end{array}
  \right\},
\end{eqnarray}
where the subscript $u$ stands for an unstable
manifold.
The conditions (\ref{eqn:cond1}) mean that the unstable manifold is
asymptotic to the hyperbolic fixed point with positive value.

We construct a formal solution $\qjtea$ of the outer equations
(\ref{eqn:outereq}) in the power series of $\ee$ as
\begin{eqnarray}
  \label{eqn:outexp}
  \qjtea & = & \ds \sum_{l=0}^{\infty}\qjal\e^{2l},
\end{eqnarray}
which we call {\it the outer expansions}.
Substituting the equations (\ref{eqn:outexp}) into the equations
(\ref{eqn:outereq}), equations are successively obtained for each
power of $\ee$;
\begin{eqnarray}
  \label{eqn:exp0}
  \O{\e^0} & : & 
  \begin{array}{rcl}
    \label{eqn:out0}
    \qjaatt & = & \qjaa - 2\qjaa^3
  \end{array},\\
  \label{eqn:outk}
  \O{\e^{2l}} & : & 
  \begin{array}{rcl}
    \qjaltt & = & ( 1 - 6\qjaa^2 )\qjal + \Fjal
  \end{array}\ (l \ge 1),
\end{eqnarray}
where $\Fjal$ are polynomials of $q_{m,00}(t), q_{m,01}(t), \cdots,
q_{m,0,l-1}(t), m = 1,2$ and their derivatives.
The boundary conditions (\ref{eqn:cond1}) are rewritten as
\begin{eqnarray}
  \label{eqn:condout1}
  \left.
    \begin{array}{rcl}
      \ds \lim_{t\to -\infty}\qjal & = & 0\\
      \qjal & > & 0
    \end{array}
  \right\}
\end{eqnarray}
for $l = 0,1,\cdots$.

From the equations (\ref{eqn:exp0}) and (\ref{eqn:condout1}) we obtain 
unperturbed solutions as
\begin{eqnarray}
  \label{eqn:outsol0}
  \qjaa & = & \sech{t-\tj}\nonumber
\end{eqnarray}
where $\tj$s are integral constants.

In this paper we restrict our attention to a 1-dimensional
sub-manifold which satisfies $\qaute = \qbute$, because it seems that
construction of the whole 2-dimensional manifold is difficult.
Hence one must put $\ta = \tb$.
Furthermore the system is autonomous; accordingly we can put $\ta =
\tb = 0$ without loss of generality.
Thus
\begin{eqnarray}
  \label{eqn:outsol}
  \qjaa & = & \sech{t}.
\end{eqnarray}

The unperturbed solutions (\ref{eqn:outsol}) possess even symmetry.
Hence let us suppose that solutions of the outer equations
(\ref{eqn:outereq}) also possess the even symmetry, i.e.,
\begin{eqnarray}
  \label{eqn:cond2}
  \qjumte & = & \qjute\jab.
\end{eqnarray}
Because the stable sub-manifold $\qjste$ is obtained as $\qjste =
\qjumte$, if one can obtain consistent solutions of the equations
(\ref{eqn:outereq}) under the conditions (\ref{eqn:cond2}), the stable
and unstable sub-manifolds coincide perfectly, i.e., homoclinic
bifurcation does not occur.
Expanding the conditions (\ref{eqn:cond2}) in the power of $\ee$ one
obtains
\begin{eqnarray}
  \label{eqn:condout2}
  q_{j,0l}(-t) & = & q_{j,0l}(t),\ l=0,1,\cdots.
\end{eqnarray}

Next, by using the equations (\ref{eqn:outsol}) we shall construct
solutions for $\O{\ee}$.
The ODEs (\ref{eqn:outk}) are linear with inhomogeneous terms
$\Fjak$.
Hence we construct homogeneous solutions first.
They are linear combinations of two independent solutions $\va$ and
$\vb$ written as
\begin{eqnarray}
  \label{eqn:outhomogs1}
  \va & = & -\sinh{t}\sechd{t},\\
  \label{eqn:outhomogs2}
  \vb & = & \ds \frac{3}{2}\sech{t} - \frac{1}{2}\cosh{t} -
  \frac{3}{2}t\,\sinh{t}\sechd{t}.
\end{eqnarray}
Note that $\va$ and $\vb$ are odd and even functions of $t$,
respectively and Wronskian $W \equiv 1$.
One can construct solutions to the equations (\ref{eqn:outk}) under the 
conditions (\ref{eqn:condout1}) and (\ref{eqn:condout2}) by using the
equations (\ref{eqn:outhomogs1}) and (\ref{eqn:outhomogs2});
\begin{eqnarray}
  \label{eqn:out1nonhomossl}
  \qjab & = & \frac{1}{3}\,\secht{t} - \frac{7+12\k}{24}\,\sech{t} +
  \frac{t}{24}\,\sinh{t}\sechd{t}.
\end{eqnarray}
In this way one can construct even solutions $\qjal\lpb$
successively.

This result leads to no homoclinic bifurcation, which is inconsistent
with the fact that the symplectic map (\ref{eqn:map1}) becomes chaotic
when $\e\ne 0$.
This indicates a breakdown of the application of the regular
perturbation method (\ref{eqn:outexp}).
The breakdown can be clearly seen near the singular points of the
outer solutions (\ref{eqn:outsol}) and (\ref{eqn:out1nonhomossl}).
We will discuss the breakdown in the next section.

\section{The breakdown of the outer expansions}
\label{sec:break}

In this section we analyze the singular points of the outer solutions
(\ref{eqn:outsol}) and (\ref{eqn:out1nonhomossl}),
extending the domain to complex $t$-plane.

The unperturbed solutions (\ref{eqn:outsol}) have first order poles at
$t = \pi i/2 + \pi i n,\ n \in {\bf Z}$.
And the solutions for $\O{\ee}$ (\ref{eqn:out1nonhomossl}) have third
order poles at the same points.
One can show that the solutions for $\O{\e^{2l}}$ have the $(2l+1)$-th 
order poles (in general, with ramification) at $t = \pi i/2 + \pi i
n,\ n \in {\bf Z}$, i.e., for example, magnitude of $\qjal$ in the
neighborhood of $t = \pi i/2$ are given by
\begin{eqnarray}
  \label{eqn:Laurant}
  \qjal & = & \ds \frac{\ajl}{\left(t-\pi i/2\right)^{2l+1}} \left( 1
    + \Osize{\left| t - \frac{\pi}{2}i \right|} \right),
\end{eqnarray}
where $\ajl$ are pure imaginary constants, for example, $\aaa = \aba =
-i$, $\aab = \abb = i/3$.
By substituting the expressions (\ref{eqn:Laurant}) into the
expansions (\ref{eqn:outexp}) one obtain
\begin{eqnarray}
  \label{eqn:outBD}
  \qjtea & = & \sum_{l=0}^{\infty}\qjal\e^{2l}\nonumber\\
  & = & \sum_{l=0}^{\infty}\frac{\ajl\e^{2l}}{\left( t - \pi i/2
    \right)^{2l+1}}\left( 1 + \Osize{\left| t - \frac{\pi}{2}i
      \right|} \right).
\end{eqnarray}
From the equations (\ref{eqn:outBD}) one can see that all the terms in
the expansions (\ref{eqn:outexp}) give contributions of the same
magnitude in $\left| t - \pi i/2 \right| \sim \e$ and that the regular 
perturbation method breaks down there.
In the next section, we analyze the region near the singular points
by blowing up there.
The region are called {\it the inner region}, whereas the region in
which the regular perturbation method is valid ($t\to -\infty$) is
called {\it the outer region}.

\section{The inner equations}
\label{sec:Inner}

In the outer region we have obtained the even solution of the
sub-manifold.
This solution is valid in asymptotic of $t\to -\infty$, but near
the singular points the asymptotic expansions (\ref{eqn:outexp}) break
down and one can't construct analytic continuation of $\qjute$ from
the left half of $t$-plane to the right one.
To construct the analytic continuation and capture odd parts of the
manifold, we shall blow up near one of the singular points, $t = \pi
i/2$ as
\begin{eqnarray}
  \label{eqn:rescaling1}
  \ds t - \frac{\pi}{2}i & = & \e z.
\end{eqnarray}
In addition, we transform $q_j(t,\e )$ to $\Phi_j(z,\e )$ in order
that the equations to solve possess no negative powers of $\e$ as
\begin{eqnarray}
  \label{eqn:rescaling2}
  \ds \e q_j(t,\e ) & = & \Phi_j(z,\e ).
\end{eqnarray}
By substituting the equations (\ref{eqn:rescaling1}) and
(\ref{eqn:rescaling2}) into the equations (\ref{eqn:mapqt}), we obtain
\begin{eqnarray}
  \label{eqn:innereq}
  \DD\Pjze & = & -2\Pjze^3 + \ee\Pjze - 2\k\e^2\Pjze\Pkze^2,
\end{eqnarray}
where $\DD\Pjze = \Pj (z+1,\e ) - 2\Pjze + \Pj (z-1,\e )$,
The second order difference equations (\ref{eqn:innereq}) are called
{\it the inner equations}.

In this paper we try to construct the leading term of $\Pjze$, hence
we restrict our attention only to the leading term of the inner
equations, i.e.,
\begin{eqnarray}
  \label{eqn:innerleadingsim}
  \DD \Pz & = & -2\Pz^3.
\end{eqnarray}
We also call the equation (\ref{eqn:innerleadingsim}) the inner
equation\footnote{In \cite{Tov2} the leading equation of the inner
  equation is called {\it the truncated inner equation}.
  Here we call the equation (\ref{eqn:innerleadingsim}) the inner
  equation for simplicity.}.

\section{The formal solution of the inner equation and Borel
  transformation}
\label{sec:InnerFormal}

One can find the formal solution to the equation
(\ref{eqn:innerleadingsim}) around $z = \infty$ in the power series of
$1/z$ as
\begin{eqnarray}
  \label{eqn:innerforsol}
  \begin{array}{rcl}
    \Paazh & = & \ds \sum_{l=0}^{\infty}\frac{\al}{z^{2l+1}}
  \end{array}
\end{eqnarray}
where $\al \defeq \aal = \abl\lab$ (see the equations
(\ref{eqn:outBD})).
Because $\Paazh$ is a solution to the equation
(\ref{eqn:innerleadingsim}) with boundary condition $\Paazh \to 0$ for
$|z| \to \infty$, the coefficients $\al,\, l=0,1,\cdots$ satisfy the
following estimation\cite{Suris94}:
\begin{eqnarray}
  \label{eqn:akEstimate}
  \al & \sim & (-1)^{l+1}\,(2l+3)!\,(2\pi)^{-2l-2}\ (l \to \infty).
\end{eqnarray}
Although the radius of convergence of the right hand side of the
equation (\ref{eqn:innerforsol}) is obviously 0, $\Paazh$ is an
asymptotic expansion of $\Puz$ for $\Re{z} < 0$ and of $\Psz$ for
$\Re{z} > 0$.
Additionary $\Paazh$ is Borel summable, i.e., the formal Borel
transformation (the inverse Laplace transformation) of $\Paazh$
\begin{eqnarray}
  \label{eqn:V_00def}
  \V & \defeq & \LapIm\Paazh\\
  \label{eqn:V_00}
  & = & \ds \sum_{l=0}^{\infty}\frac{\al}{(2l)!}p^{2l}\nonumber
\end{eqnarray}
has a finite radius ($= 2\pi$) of convergence and the solution
$\Phi(z) = \Lap\V$ is known as the Borel sum of $\Paazh$.

We hence apply the formal Borel transformation to the inner
equation (\ref{eqn:innerleadingsim}) and obtain
\begin{eqnarray}
  \label{eqn:BorelTra1}
  2(\cosh{p} - 1)\V & = & -2\ [\ V*V*V\ ](p),
\end{eqnarray}
where $\V$ has been defined as the equation (\ref{eqn:V_00def}) and
convolution $[\,V*W*Y\,](p)$ is defined as
\begin{eqnarray}
  \label{eqn:convolution}
  [\ V*W*Y\ ](p) & \defeq & [\ [\ V*W\ ]*Y\ ](p)\nonumber\\
  & = & [\ V*[\ W*Y\ ]\ ](p)\nonumber\\
  & = & \ds \int_{0}^{p}V(\tau )\int_{0}^{p-\tau}W(p-\tau -\lambda
  )Y(\lambda )d\lambda d\tau.\nonumber
\end{eqnarray}

The solution to the integral equation (\ref{eqn:BorelTra1}) possesses
poles only at $p=2\pi ik,\ k\in\Z\verb+\+\{0\}$\cite{TovinvLap}.
Note that these singular points are determined by the kinetic terms of
the generating function, hence in many physical problems this type of
singular points appears.

One can define the unstable sub-manifold $\Puz$ and the stable one
$\Psz$ by using the Laplace transformation as
\begin{eqnarray}
  \label{eqn:Laplace1}
  \Psz & = & \ds \int_{0}^{\infty}e^{-pz}\V dp\ ,\ |\arg{z}|<\pi,\\
  \label{eqn:Laplace2}
  \Puz & = & \ds \int_{0}^{-\infty}e^{-pz}\V dp\ ,\ -2\pi <\arg{z}<0,
\end{eqnarray}
respectively.
Now let the difference between $\Psz$ and $\Puz$ be $\Ppm$ when
$\pm\Im{z} > 0$, i.e.,
\begin{eqnarray}
  \label{eqn:imPhi_u}
  \Ppm & \defeq & \Psz - \Puz\ ( \pm\Im{z} > 0 ).
\end{eqnarray}
From the equations (\ref{eqn:Laplace1}) and (\ref{eqn:Laplace2}) one
obtains
\begin{eqnarray}
  \label{eqn:imPhi_uvsP_m}
  i\ \Im{\Pu(-i\zeta )} & = & -\frac{1}{2}\Pm(-i\zeta ),
\end{eqnarray}
where $\zeta$ is an arbitrary positive real number (see Appendix
\ref{sec:InnerIm}).

\begin{figure}[ht]
  \begin{center}
    \leavevmode
    \epsfxsize=8cm
    \epsfbox{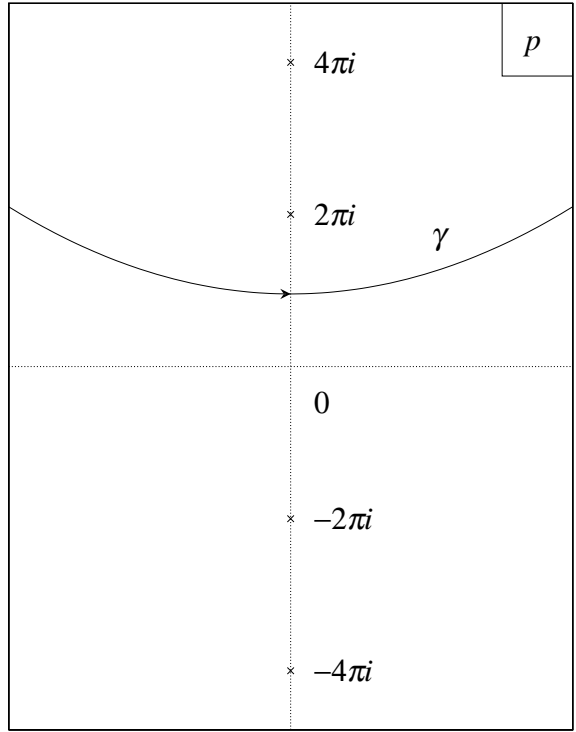}
    \caption{The integral path $\gamma$ on complex $p$ plane is
      shown. One can take $\gamma$ as enclosing the positive imaginary
      semi-axis.}
    \label{fig:intpath}
  \end{center}
\end{figure}

Furthermore from the equations (\ref{eqn:Laplace1}),
(\ref{eqn:Laplace2}) and (\ref{eqn:imPhi_u}) one obtains
\begin{eqnarray}
  \label{eqn:Phi_mIntRep}
  \Pmz & = & \int_{\gamma}\exp{-pz}\V dp\ ( -\pi < \arg{z} < 0 ).
\end{eqnarray}
The integral path $\gamma$ is shown in figure \ref{fig:intpath}.
$\Pm(z)$ is given as the sum of the residues of the poles on 
the positive imaginary semi-axis, i.e.,
\begin{eqnarray}
  \label{eqn:residue}
  \Pmz & = & 2\pi i\sum_{k=1}^{\infty}\Res{\exp{-pz}\V}{p=2\pi ik}.
\end{eqnarray}

Because of $|\exp{-2\pi iz}| \gg |\exp{-4\pi iz}| \gg \cdots$ in
$\Im{z} < 0$, we shall only take account of the first term of the
right hand side of the equation (\ref{eqn:residue}), i.e.,
\begin{eqnarray}
  \label{eqn:residue2}
  \Pmz & \sim & 2\pi i\ \Res{\exp{-pz}\V}{p=2\pi i}. \nonumber
\end{eqnarray}

From the equation (\ref{eqn:residue}) one can realize that $\Puz$ and
$\Psz$ possess the term in proportion to $\exp{-2\pi ikz}$, hence we
look for solutions to the equation (\ref{eqn:innerleadingsim}) with
the form
\begin{eqnarray}
  \label{eqn:innerexp}
  \Pc (z) & = & \sum_{l=0}^{\infty}\Palzh\exp{-2\pi ilz}\\
  & = & \Paazh + \Pabzh\exp{-2\pi iz} + \cdots.\nonumber
\end{eqnarray}
By substituting the equation (\ref{eqn:innerexp}) into the equation
(\ref{eqn:innerleadingsim}), equations are successively obtained for
each power of $\exp{-2\pi iz}$;
\begin{eqnarray}
  \label{eqn:innerexpexp0}
  \O{\exp{0}} & : &
  \begin{array}{rcl}\DD\Paazh & = & -2\Paazh^3,
  \end{array}\\
  \label{eqn:innerexpexp1}
  \O{\exp{-2\pi iz}} & : &
  \begin{array}{rcl}\DD\Pabzh & = &
    -6\Paazh^2\Pabzh,
  \end{array}\\
  & \cdots. &\nonumber
\end{eqnarray}

The formal solution to the equation (\ref{eqn:innerexpexp0}) has been
obtained as the equation (\ref{eqn:innerforsol}).
In order to find the leading term of a solution to the equation
(\ref{eqn:innerexpexp1}), we replace $\Paazh$ by its leading term
$-i/z$.
This yields
\begin{eqnarray}
  \label{eqn:inner1}
  \DD\Pabzh & = & \frac{6}{z^2}\Pabzh.
\end{eqnarray}
The equation (\ref{eqn:inner1}) possesses two formal solutions $\Pbe
,\ \Pbo$ whose leading terms are $z^{-2}, z^3$, respectively.
Because the odd solution $\Pbo$ dominates the even solution $\Pbe$, we
take account of only $\Pbo$, i.e., we replace $\Pabzh$ as $\Pbo$.

One can continue the process and obtain the formal solutions of $\Plh
\lpb$.
In this paper, however, we restrict our attention only to the first
two terms of the equation (\ref{eqn:innerexp}), $\Paazh$ and $\Pabzh$.

The odd solution $\Pbo$ possesses third order poles at $z = \infty$.
Hence it is easily proved that $\V$ possesses fourth order poles at $p
= 2\pi ik,\ k\in\Z$ by properties of Laplace transformation, i.e.,
\begin{eqnarray}
  \label{eqn:LapImPjh}
  \V & = & \LapIm\Paazh \nonumber \\
  \label{eqn:LapImPjh2}
  & \sim & \frac{\dk}{(p-2\pi ik)^4} + \Osize{\frac{1}{|(p-2\pi
      ik)^3|}},\ \dk :\const ,\ k\in\Z.
\end{eqnarray}
From the equation (\ref{eqn:LapImPjh2}) we can obtain $\Pmz$ as
\begin{eqnarray}
  \label{eqn:residue22}
  \Pmz & = & 2\pi i\lim_{z\to 2\pi i}\frac{1}{3!}\frac{d^3}{dp^3}[(p-2\pi
  i)^4\exp{-pz}\V] \nonumber \\
  & = & cz^3\exp{-2\pi iz}, \nonumber
\end{eqnarray}
where $c$ ( $=-2\pi id_1/3!$ ) is a {\it Stokes constant} to be
evaluated.
There have been no analytical methods to evaluate the Stokes constant.
However it is possible to estimate numerically by using the same
algorithm as used in \cite{Pomeau,Tov1,Tov2,NakaHama}.
The result of our calculation shows $c \sim -796$.

By using the equation (\ref{eqn:imPhi_uvsP_m}) one can let the
coefficient of $bz^3\exp{-2\pi iz}$, which is the leading term of
$\Pjb$, as
\begin{eqnarray}
  \label{eqn:prefactor}
  b & = & \left\{
    \begin{array}{rcl}
      0 & , & \ds -\pi < \arg{z} < -\frac{\pi}{2}\\
      \ds -\frac{c}{2} & , & \ds \arg{z} = -\frac{\pi}{2}\\
      -c & , & \ds -\frac{\pi}{2} < \arg{z} < 0
    \end{array},
  \right.
\end{eqnarray}
approximately.

\section{Matching of the solutions}
\label{sec:Matching}

In this section we match the inner solutions to the outer solutions
and construct the unstable sub-manifold which is valid on the right
half of $t$-plane.

First we give the exponential expansions of the outer variables.
Following the inner expansion (\ref{eqn:innerexp}), the refined outer
expansions are given by
\begin{eqnarray}
  \label{eqn:outexpexp}
  \qjteci & = & \qjtea + \qjteb\exp{-\frac{2\pi i}{\e}t} +
  \qjtec\exp{-\frac{4\pi i}{\e}t} + \cdots.
\end{eqnarray}
Note that the exponential expansion (\ref{eqn:outexpexp}) is singular
with respect to $\e$, i.e., $\exp{-\frac{2\pi i}{\e}t}$ has essential
singularity at $\e = 0$.

The leading term of the expansion (\ref{eqn:outexpexp}) has been
obtained in section \ref{sec:outer} as
\begin{eqnarray}
  \label{eqn:outersolagain}
  \qjtea & = & \ds \qjaa + \e^2\qjab + O(\e^4), \nonumber
\end{eqnarray}
where
\begin{eqnarray}
  \label{eqn:outj00ag}
  \qjaa & = & \sech{t}, \\
  \label{eqn:outj01ag}
  \qjab & = & \ds \frac{1}{3}\,\secht{t} - \frac{7+12\k}{24}\,\sech{t}
  + \frac{1}{24}t\,\sinh{t}\sechd{t}.
\end{eqnarray}

Next we construct the first exponential terms $\qjteb$, which
correspond to $\O{\exp{-\frac{2\pi i}{\e}t}}$.
Substituting the expansions (\ref{eqn:outexpexp}) into the equations
(\ref{eqn:mapqt}) gives 
\begin{eqnarray}
  \label{eqn:outexpexp1}
  \DDe\qjteb & = & ( 1 - 6\qjtea^2 )\qjteb - 2\e^2\k\qjtea\qktea^2.
  \nonumber
\end{eqnarray}

To estimate the leading order of $\qjtba$, which is the leading term
of $\qjteb$, we replace $\qjtea$ with their leading term $\qjaa$.
By these replacements, we obtain the differential equations of
$\qjtba$ as
\begin{eqnarray}
  \label{eqn:outexpexp10}
  \frac{d^2\qjtba}{dt^2} & = & \ds ( 1 - 6\qjaa^2 )\qjtba.
\end{eqnarray}
The general solutions of the equations (\ref{eqn:outexpexp10}) are
given as
\begin{eqnarray}
  \label{eqn:q_j10}
  \qjtba & = & c_{j,1}\va + c_{j,2}\vb, \nonumber
\end{eqnarray}
where $c_{j,1}$ and $c_{j,2}$ are arbitrary constants, and $v_1(t)$ and 
$v_2(t)$ are given by the equations (\ref{eqn:outhomogs1}) and
(\ref{eqn:outhomogs2}).

We rewrite functions $\va$ and $\vb$ in the neighborhood of the
singular point $t = \pi i/2$.
If we introduce $\d = t - \pi i/2$, the equations
(\ref{eqn:outhomogs1}) and (\ref{eqn:outhomogs2}) imply
\begin{eqnarray}
  \label{eqn:v_adelta}
  \va & = & i\frac{\cosh{\d}}{\sinhd{\d}},\\
  \label{eqn:v_bdelta}
  \vb & = & \frac{3}{4}\pi i\va \nonumber \\
  & & -
  \frac{1}{2}i\frac{1}{\sinhd{\d}}[\,3\sinh{\d} + \sinht{\d} -
  3\d\cosh{\d} \, ].
\end{eqnarray}
One can see that the odd function $\va$ becomes an even function of
$\d$ and the even function $\vb$ has both odd and even parts.
The leading term of $\va$ around $\d = 0$ is of the order $\O{\d^{-2}}
= \O{\e^{-2}z^{-2}}$.
It corresponds to the solution $\Pbe$ of equation (\ref{eqn:inner1}),
which is sub-dominant to the other solution $\Pbo$.
The odd part of $\vb$, which is in the square bracket, has the Taylor
expansion as
\begin{eqnarray}
  \label{eqn:v_bTaylor}
  \lefteqn{3\sinh{\d} + \sinht{\d} - 3\d\cosh{\d}} \nonumber \\
  & = & 0\cdot\d + 0\cdot\d^3 + \frac{2}{5}\d^5 + \O{\d^7}. \nonumber
\end{eqnarray}
The coefficient of the square bracket is of the order $\O{\d^{-2}}$.
These show that the odd part of the equation (\ref{eqn:v_bdelta}) is
of the order $\O{\d^3}\,(=\O{\e^3z^3})$, corresponding to $\Pbo$.
We shall therefore match $\qjtba$ to $\Pabzh \equiv \Pbo$, i.e.,
$\qjtba$ is given by the odd part of $\vb$ as
\begin{eqnarray}
  \label{eqn:q_j10leading}
  \qjtba & = & c_{j,2}\left[\vb - \frac{3}{4}\pi i\va \right]\\
  & = & -\frac{1}{5}c_{j,2}i\d^3(1+\O{\d^2})\nonumber\\
  & = & -\frac{1}{5}c_{j,2}i\e^3z^3(1+\O{\d^2}). \nonumber
\end{eqnarray}
Furthermore the discussion in section \ref{sec:InnerFormal} shows
\begin{eqnarray}
  \label{eqn:P_j11oddleading}
  i\ \Im{\Puz} & \sim & bz^3\exp{-2\pi iz} \nonumber
\end{eqnarray}
where $b$ has been determined as the equation (\ref{eqn:prefactor}).
The matching $\qjtba$ and $\Pbo$ for t approaching $\pi i/2$ with
$\arg{z} = -\pi /2$ is given below:
\begin{eqnarray}
  \label{eqn:matching0}
  \e\qjtba\Exp{-\frac{2\pi it}{\e}} & = & -\frac{1}{2}cz^3\exp{-2\pi
    iz}, \nonumber
\end{eqnarray}
hence, 
\begin{eqnarray}
  \label{eqn:c_2}
  c_{j,2} & = & -\frac{5}{2}i\frac{c}{\e^4}\Exp{-\frac{\pi^2}{\e}},
\end{eqnarray}
where $c\in\R$, therefore, $c_{j,2}\in\iR$.
Another constant $c_{j,1}$ is given as
\begin{eqnarray}
  \label{eqn:c_1}
  c_{j,1} & = & \frac{3}{4}\pi ic_{j,2} \nonumber \\
  & = & \frac{15}{8}\pi\frac{c}{\e^4}\Exp{-\frac{\pi^2}{\e}} \nonumber
\end{eqnarray}
(see equation (\ref{eqn:q_j10leading}) and (\ref{eqn:c_2})).

Finally, the leading term of the imaginary part of the sub-manifold
$\qjtba$ is given as
\begin{eqnarray}
  \label{eqn:q_jbafinal}
  \qjtba & = & c_{j,1}\va + c_{j,2}\vb\nonumber\\
  & = &
  -\frac{1}{2}ic_{j,2}\frac{1}{\sinhd{\d}}
  [3\sinh{\d}+\sinht{\d}-3\d\cosh{\d}] \nonumber \\
  & = &
  \frac{c_{j,2}}{2} \left[3(t-\frac{\pi}{2}i)\sinh{t}\sechd{t} - 3\sech{t} 
    + \cosh{t}\right] \nonumber.
\end{eqnarray}

\section{Results}
\label{sec:result}

We have constructed the solutions which are valid in the neighborhood
of the singular point $t = \pi i/2$.
The same arguments, applied to the complex conjugated singular point
$t = - \pi i/2$, yield the complex conjugated solutions.
We take the two singular points, which are nearest to the real axis,
into consideration and restore the obtained solutions to real
functions.
Taking the average of these two solutions, we obtain
\begin{eqnarray}
  \label{eqn:mean}
  \qjteci & = & \qjtea +
  \Re{\left[\sum_{l=1}^{\infty}\qjtel\Exp{-\frac{2\pi
          il}{\e}t}\right]}. \nonumber
\end{eqnarray}
In this paper we approximate the sub-manifold by taking the term $l =
1$.
This term is written as
\begin{eqnarray}
  \label{eqn:meanleading}
  \Re{\left[\qjteb\Exp{-\frac{2\pi it}{\e}}\right]} & \sim &
  \Re{\left[\qjtba\Exp{-\frac{2\pi it}{\e}}\right]} \nonumber \\
  & = & c_1\va\cosn{\frac{2\pi t}{\e}} - ic_2\vb\sinn{\frac{2\pi
      t}{e}} \nonumber \\
  \label{eqn:MBdef}
  & = & M(t,\e )B(t,\e ) \nonumber
\end{eqnarray}
where
\begin{eqnarray}
  \label{eqn:Mdef}
  M(t,\e ) & = & -S(t,\e
  )\frac{5}{2}\frac{c}{\e^4}\Exp{-\frac{\pi^2}{\e}},\\
  \label{eqn:Bdef}
  B(t,\e ) & = & \frac{3}{4}\pi\va\cosn{\frac{2\pi t}{\e}} -
  \vb\sinn{\frac{2\pi t}{\e}}.
\end{eqnarray}
The switching function $S(t,\e )$, which stands for Stokes phenomenon, 
is approximately given as
\begin{eqnarray}
  \label{eqn:Sdef}
  S(t,\e ) & = & \left\{
    \begin{array}{cc}
      0 & ( t < 0 )\\
      1 & ( t = 0 )\\
      2 & ( t > 0 )
    \end{array}
  \right.
\end{eqnarray}
(see the equations (\ref{eqn:prefactor}) and (\ref{eqn:Mdef})).

By using these functions, the parametrization of the unstable
sub-manifold is
\begin{eqnarray}
  \label{eqn:q_j1u}
  \qjteut & = & \qjaa + \ee\qjab + M(t,\e )B(t,\e ),
\end{eqnarray}
and the momentum variables $\pjteut$ is given as
\begin{eqnarray}
  \label{eqn:pjte}
  \pjteut & = & \frac{\qjteut-\qjtmeut}{\e}
\end{eqnarray}
(see the equation (\ref{eqn:map1})).
On the other hand, the parametrization of the stable sub-manifold is
given as $\qjtest = \qjteumt$ and $\pjtest = -\tilde{p}_{j,u}(-t+\e
,\e )$.

By using the equations (\ref{eqn:pjte}) one can observe the manifold
on the phase space.
However it is impossible to display a 4-dimensional phase space
graphically, hence we observe its projections to $(\qa ,\pa )$ and
$(\qb ,\pb )$ plane.
From equations (\ref{eqn:q_j1u}), (\ref{eqn:outj00ag}) and
(\ref{eqn:outj01ag}) the projections to $(\qa ,\pa )$ and $(\qb ,\pb
)$ plane are same.
\begin{figure}[hbtp]
  \begin{center}
    \epsfxsize=7cm
    \label{pic:sep1}
    \subfigure[]{\epsffile{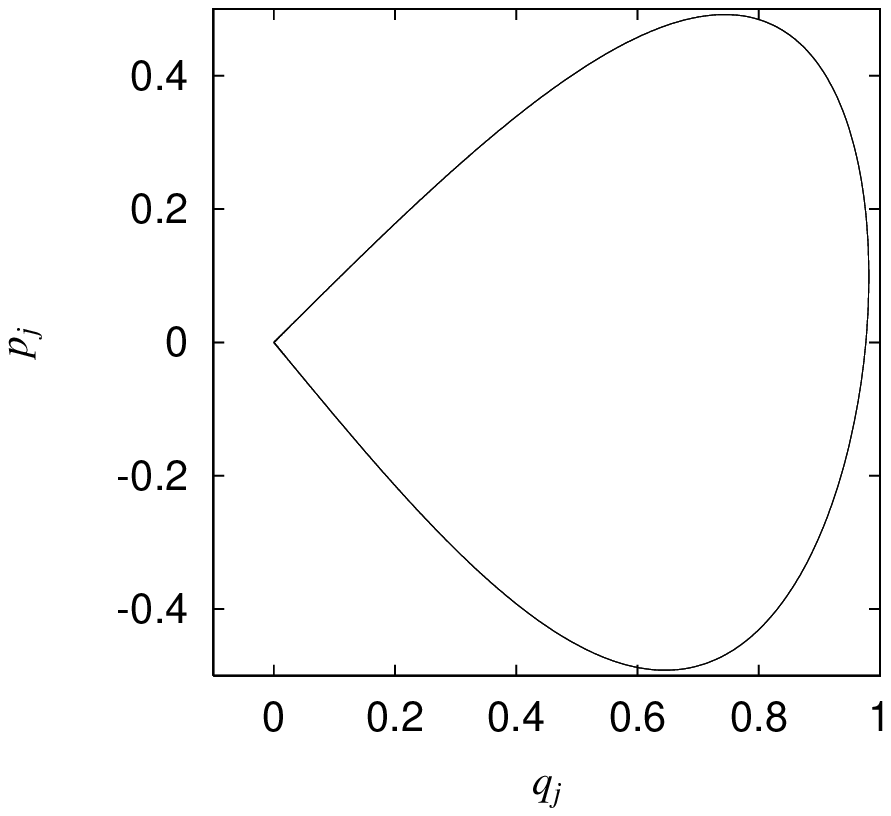}}
    \epsfxsize=7cm
    \subfigure[]{\epsffile{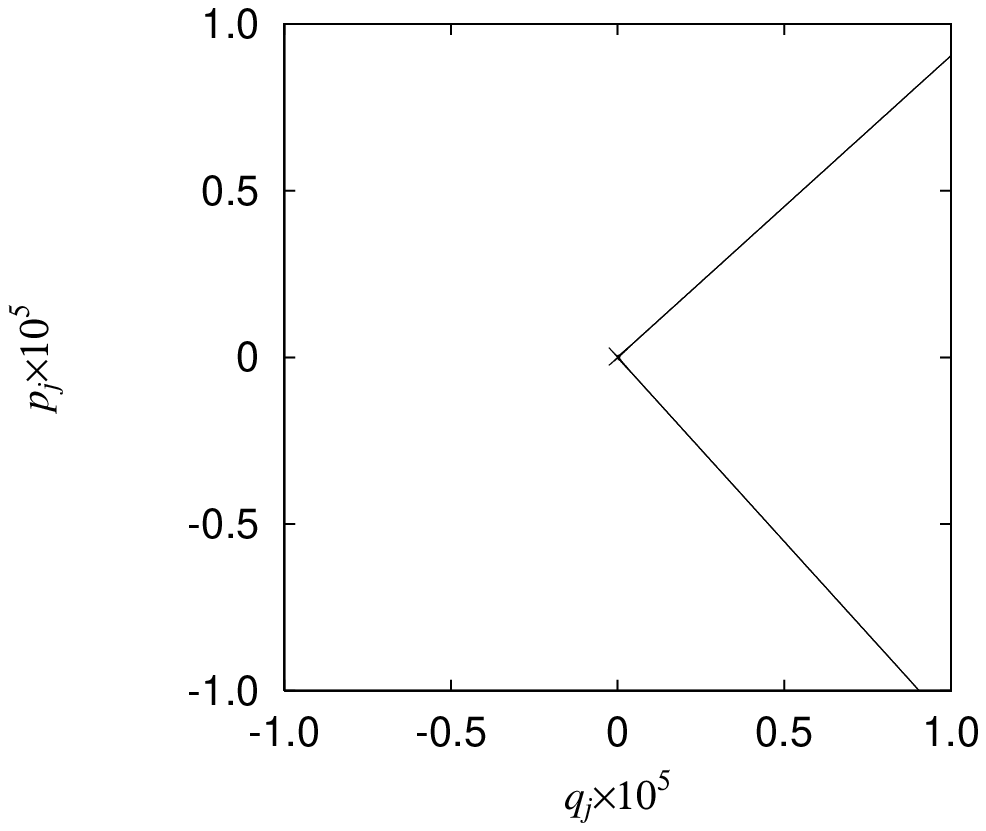}}
    \epsfxsize=7cm
    \subfigure[]{\epsffile{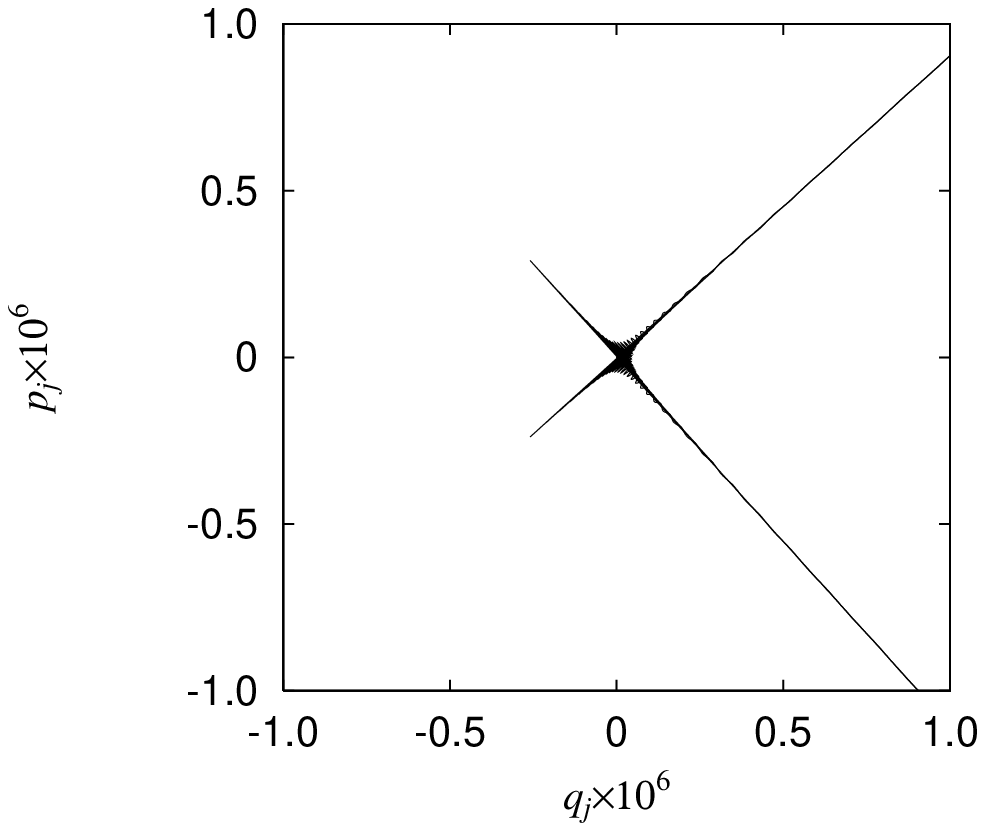}}
    \epsfxsize=7cm
    \subfigure[]{\epsffile{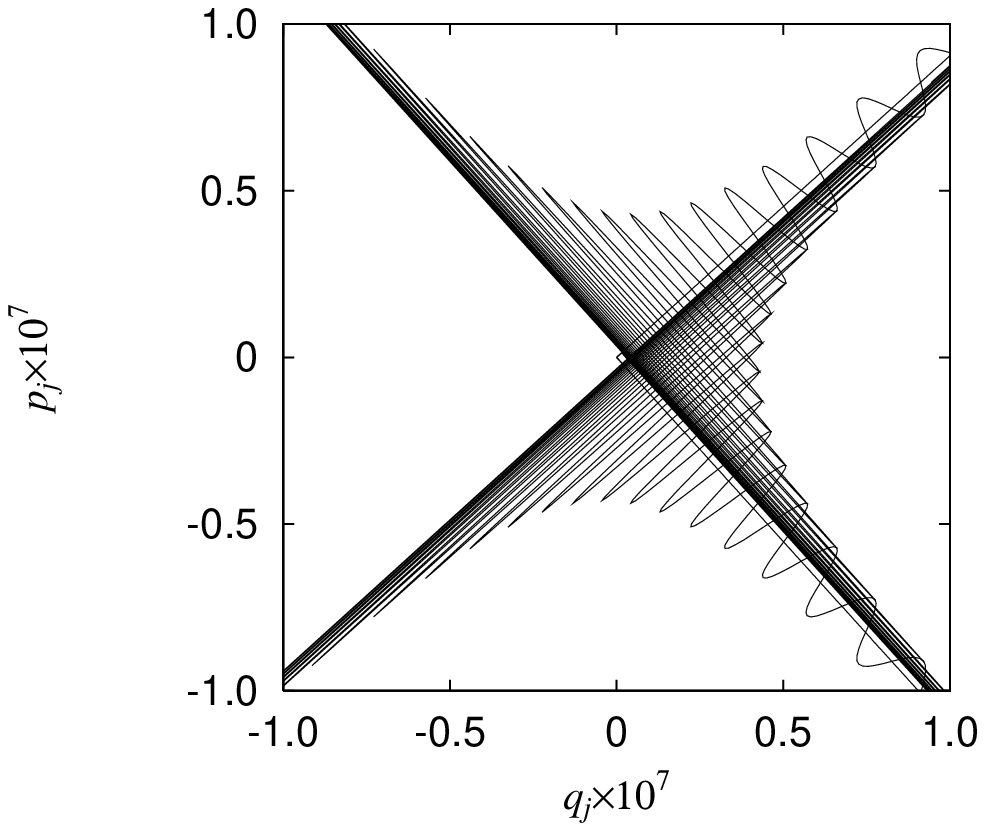}}
  \end{center}
  \caption{The functional approximations of the stable and unstable
    sub-manifolds projected on $(\qj ,\pj )$ plane. The parameters are
    $\e = 0.2$, $\k = 1.0$.}
  \label{fig:Sep2e30UM}
\end{figure}
Figure \ref{fig:Sep2e30UM},(a) shows that global shapes of the stable and unstable sub-manifolds.
In figure \ref{fig:Sep2e30UM},(b)--(d) we magnify the neighborhood of
the fixed point.
One can clearly see the homoclinic structure from figure
\ref{fig:Sep2e30UM},(d).

Next we compare the approximate solution (\ref{eqn:q_j1u}) and
(\ref{eqn:pjte}) with numerical trajectories.
\begin{figure}[hbtp]
  \begin{center}
    \leavevmode
    \epsfxsize=11cm
    \epsfbox{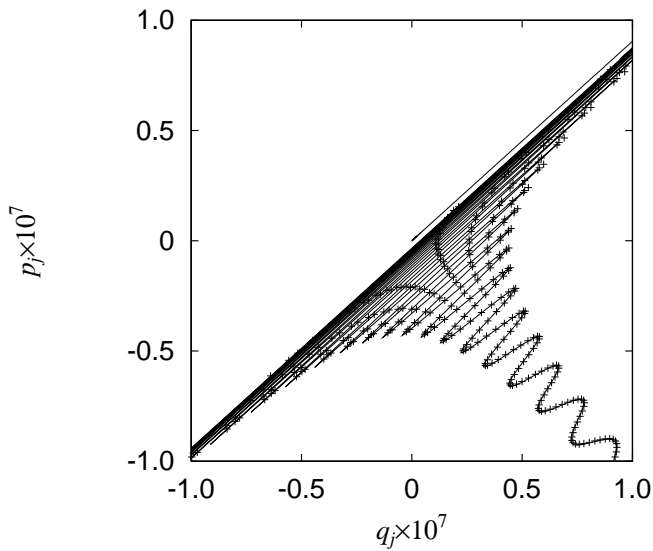}
    \caption{The comparison between the approximate solution of the
      unstable sub-manifold and 20 numerical trajectories. The
      parameters are same as figure \ref{fig:Sep2e30UM}. The initial
      values of the numerical trajectories are given on the solution
      between $t = 14.0$ and $t = 14.2$ with equal intervals.}
    \label{fig:e20k10}
  \end{center}
\end{figure}
20 numerical trajectories are plotted on figure \ref{fig:e20k10}.
This figure shows that the approximate solution agree well with the
numerical trajectories.
The initial values of the numerical trajectories are given on the
approximate solutions between $t = 14.0$ and $t = 14.2$ with equal
intervals, i.e., $t = 14.0$, $t = 14.005$, $\cdots$, $t = 14.195$.
In other values of $\e$ or $\k$ also one can show that the approximate 
solution agree well with numerical solutions.

\section{Summary and discussions}
\label{sec:Summary}

We extended the method {\it ``asymptotic expansions beyond all
  orders''} for the 4-dimensional symplectic mapping (\ref{eqn:map1})
straightforwardly and successfully obtained the explicit functional
approximation of the 1-dimensional sub-manifolds of the stable and
unstable manifolds (\ref{eqn:q_j1u}).
These asymptotic solutions well approximate numerical solutions in the
neighborhood of the hyperbolic fixed point, which is important to
understand phase space's structure of Hamiltonian systems.
We therefore consider that by this extension one can analyze global
structure of phase space of Hamiltonian systems with three degree of
freedom.

The oscillating parts $M(t,\e )$ and $B(t,\e )$ do not contain the
coupling constant $\k$ (see the equations (\ref{eqn:Mdef}) and
(\ref{eqn:Bdef})).
This indicates that the homoclinic structure is a direct product of
2-dimensional symplectic mappings.
In other words, the model we analyzed shows only quantitative
characteristic properties of high-dimensionality.
This is based on the fact that the inner equation
(\ref{eqn:innerleadingsim}) is decoupled.
From the order of the singular points of the outer solutions, we find
the condition that the coupling term dominates the nonlinear terms.
The condition is written as
\begin{eqnarray}
  \label{eqn:condcouple}
  \left\{
    \begin{array}{rclcl}
      \a_1 + \a_2 & < & \gamma + 4 & \cdots & {\rm decouple} \\
      \a_1 + \a_2 & > & \gamma + 4 & \cdots & {\rm coupling\ term\
        dominates}
    \end{array}
  \right.
\end{eqnarray}
(see the map (\ref{eqn:map0}) and appendix \ref{sec:condition}).
When the second condition of the conditions (\ref{eqn:condcouple}) is
satisfied, the coupling term dominantly contributes the inner
equations and another type of rescaling (see equations
(\ref{eqn:rescaling1}) and (\ref{eqn:rescaling2})) is required.
Hence we consider that the models whose coupling terms satisfy the
second condition make not only quantitative but also qualitative
differences.
However it is impossible to treat such models with the same way.
We would like to improve this method in order to extend to such models 
in a future work.

In the present work we have restricted our attention to the particular 
1-dimensional sub-manifolds of the stable and unstable manifolds.
And we have shown that the sub-manifolds oscillate in the vicinity of
the origin.
Because the unstable sub-manifold is contained in the unstable
manifold, it is clear that the unstable manifold itself oscillates.
We now try to construct explicit functional approximations of the
whole 2-dimensional stable and unstable manifold\cite{HNK}.

In the present model there exists a 4-dimensional fully hyperbolic
fixed point (a direct product of two 2-dimensional hyperbolic fixed
points).
In a future work we will extend this method to 4-dimensional
symplectic mappings with a hyperbolic$\times$elliptic type fixed
point.
\vspace{1em}

\section*{Acknowledgements}

We would like to thank K. Nozaki for helpful discussions and
suggestions.

\appendix

\section{The imaginary part of $\Puz$}
\label{sec:InnerIm}

In this appendix we prove equation (\ref{eqn:imPhi_uvsP_m}).
From equations (\ref{eqn:Laplace1}) and (\ref{eqn:Laplace2}) we obtain 
\begin{eqnarray}
  \Phi_u(-z) & = & \ds \int_{0}^{-\infty}e^{pz}\V dp\nonumber\\
  & = & \ds \int_{0}^{\infty}e^{-pz}V(-p)(-dp)\nonumber\\
  \label{eqn:P_uvsP_s}
  & = & -\Psz, \nonumber
\end{eqnarray}
and
\begin{eqnarray}
  \label{eqn:Phi_svscc}
  \Phi_s(\zb ) & = & - \Pbar_s(z). \nonumber
\end{eqnarray}

Assume $\zeta \in \R$.
One can easily prove that
\begin{eqnarray}
  \label{eqn:Phi_u(-izeta)vsPhi_s}
  \begin{array}{ccccc}
    \Phi_u(-i\zeta ) & = & -\Phi_s(i\zeta ) & = & \Pbar_s(-i\zeta
    ). \nonumber
  \end{array}
\end{eqnarray}
Furthermore assume $\zeta > 0$, and $\Ps$ and $\Pu$ are real-valued on 
the negative imaginary semi axis, one can obtain
\begin{eqnarray}
  \label{eqn:phi_uphi_s}
  \begin{array}{ccccc}
    \Phi_u(-i\zeta ) & = & \Pbar_s(-i\zeta ) & = & \Phi_s(-i\zeta ),
  \end{array}
\end{eqnarray}
\begin{eqnarray}
  \label{eqn:phi_uphi_s2}
  \begin{array}{ccccccc}
    \Phi_u(i\zeta ) & = & -\Phi_s(-i\zeta ) & = & -\Pbar_s(-i\zeta ) & = 
    & \Phi_s(i\zeta ).
  \end{array}
\end{eqnarray}
By using equation (\ref{eqn:phi_uphi_s}) and equation
(\ref{eqn:phi_uphi_s2}), we can say that
\begin{eqnarray}
  \label{eqn:Phi_ueqPhi_s}
  \Psz & \equiv & \Puz \ (z\in i\R ). \nonumber
\end{eqnarray}
Hence $\Re{\Psz} = \Re{\Puz}$ on the imaginary axis: i.e.,
\begin{eqnarray}
  \label{eqn:P_pm}
  \Ppm & \in & i\R\ (z\in i\R ). \nonumber
\end{eqnarray}
Particularly when $\zeta > 0$, from equation (\ref{eqn:phi_uphi_s})
one can obtain
\begin{eqnarray}
  \label{eqn:Phi_m}
  \Pm (-i\zeta ) & = & \Ps (-i\zeta ) - \Pu (-i\zeta )\nonumber\\
  & = & -2i\ \Im{\Pu} (-i\zeta ). \nonumber
\end{eqnarray}
Thus
\begin{eqnarray}
  \label{eqn:imPhi_uvsP_m2}
  i\ \Im{\Pu(-i\zeta )} & = & -\frac{1}{2}\Pm(-i\zeta ). \nonumber
\end{eqnarray}

\section{The order of the coupling term}
\label{sec:condition}

In this appendix we study the order of the coupling terms in the
map (\ref{eqn:map0}).
The coupling term is generated by $\e^\gamma\tilde{J} =
\e^\gamma\qan^{\a_1}\qbn^{\a_2},\, \a_j\in\N,\, j=1,2$.

Put $A = \aa + \ab$.
First let us suppose $\gamma = 2$.
The outer solutions of $\O{\e^2}$ are written as
\begin{eqnarray}
  \label{eqn:outer01cond}
  \qjab & = & \frac{1}{3}\secht{t} - \frac{7}{24}\sech{t} \nonumber \\
  & & + \frac{t}{24}\ \sinh{t}\sechd{t} + \k \aj C(t;A), \\
  C(t;A) & = & 
  \left\{
    \begin{array}{ll}
      \ds \frac{1}{3}\sechd{t} & (A = 3) \vspace{0.5em}\\
      \ds \frac{1}{A} \frac{(A-4)!!}{(A-5)!!} & \vspace{0.2em}\\
      \ds \times \sum_{k=1}^{M} \frac{(A-5-2k)!!}{(A-2-2k)!!}
      F(t;-A+1+2k) & (A \ge 4)
    \end{array}
  \right.
\end{eqnarray}
where $(2n)!! = 2n \cdot (2n-2) \cdots 2,\, (2n-1)!! = (2n-1) \cdot
(2n-3) \cdots 1, n\in\N,\, 0!! = (-1)!! = 1,\, (-2)!! = (-3)!! = -1$,
and $F(t;n), n\in Z$ is defined as
\begin{eqnarray}
  \label{eqn:sfunc}
  F(t;n) & = & 
  \left\{
    \begin{array}{ll}
      {\rm cosh}^n(t) & (n \ge 1)\\
      \sinh{t}\sechd{t}{\rm gd}(t) & (n = 0)
    \end{array}
  \right.,\\
  {\rm gd}(t) & = & {\rm tan}^{-1}(\sinh{t}) \nonumber
\end{eqnarray}
and
\begin{eqnarray}
  \label{eqn:Mcdef}
  M & \defeq & \left[ \frac{A}{2} \right] - 1\\
  & = & 
  \left\{
    \begin{array}{lcl}
      \ds \frac{A}{2} - 1 & \cdots & A:{\rm even}\vspace{0.5em}\\
      \ds \frac{A+1}{2} - 2 & \cdots & A:{\rm odd}
    \end{array}\nonumber
  \right..
\end{eqnarray}

From equations (\ref{eqn:outer01cond})--(\ref{eqn:Mcdef}) one sees
that when $A \ge 6$ the outer solutions $\qjab\jab$ possess
singularities of the order $A - 3 \, (\ge 3)$ at $t = \pi i/2 + \pi i
k,\ k \in {\bf Z}$ (in general, with ramification).
On the other hand, when $A < 6$, the order of the singularities is $3$.
One can easily prove that the outer solutions of $\O{\e^{2k}}$ possess 
singularities of the order $k(A-4) + 1$ or $2k + 1$ when $A \ge 6$ or
$A < 6$, respectively.
Hence if $A \ge 6$, the magnitude of $\qjak$ in the neighborhood
of $t = \pi i/2$ are given by
\begin{eqnarray}
  \label{eqn:LaurantCond}
  \begin{array}{rcl}
    \ds \qjak & = & \ds \frac{\ajk}{\left(t-\pi i/2\right)^{k(A - 4) +
        1}} \left( 1 + \Osize{\left| t - \frac{\pi}{2}i \right|}
    \right), \\
    \ds \ajk & \in & {\bf C}\jab\kpaa.
  \end{array}
\end{eqnarray}
From equations (\ref{eqn:LaurantCond}) one can realize that all the
terms in the formal expansions (\ref{eqn:outexp}) give contributions
of the same order in $\left| t - \pi i/2 \right| \sim \e^\delta,
\delta = \frac{2}{A-4}$.
We must therefore blow up near the one of the singular points, $t =
\pi i/2$ as
\begin{eqnarray}
  \label{eqn:rescaling3}
  \ds t - \frac{\pi}{2}i & = & \e^\delta z \nonumber
\end{eqnarray}
(compare equation (\ref{eqn:rescaling1}) and the discussions in
section \ref{sec:break}).
Note that if $A > 6$, then $\delta < 1$, hence $\e^\delta > \e$, 
that is, the inner region is wider with $A > 6$ than with $A \le 6$.

Now we show that when $\gamma = 2$ and $A > 6$, the rescaling factor
is different from that of $A \le 6$.
Hence we call the models with $A > 6$ or $A \le 6$ {\it strongly} or
{\it weakly coupled model}, respectively.
In the same way, one can prove that the condition with which the
rescaling factor is different from 2-dimensional mappings (storongly
coupled model) is written as
\begin{eqnarray}
  \label{eqn:condcond}
  \a_1 + \a_2 & > & \gamma + 4
\end{eqnarray}
for general $\gamma\in\N$.




\end{document}